\documentclass{article}

\usepackage{arxiv}

\usepackage[utf8]{inputenc} 
\usepackage[T1]{fontenc}    
\usepackage[hypertexnames=false]{hyperref}      
\usepackage{url}            
\usepackage{booktabs}       
\usepackage{amsfonts}       
\usepackage{nicefrac}       
\usepackage{microtype}      
\usepackage{graphicx}
\usepackage{doi}
\usepackage[sort&compress,square,comma,authoryear]{natbib}
\usepackage{multirow}
\usepackage{tabularx}
\usepackage{xurl}
\usepackage{appendix}

\graphicspath{ {./images/} }

\title{Personalized communication strategies: Towards a new debtor typology framework}

\author{ {Minou Ghaffari}\\
	PAIR Finance GmbH\\
	Berlin, Germany\\
	Psychology School, Faculty of Business \& Media\\
	Fresenius University of Applied Sciences\\
	Hamburg, Germany\\
	\texttt{minou.ghaffari@ext.hs-fresenius.de} \\
	\And
	{Maxime Kaniewicz} \\
	PAIR Finance GmbH\\
	Berlin, Germany\\
	\texttt{maxime.kaniewicz@pairfinance.com} \\
	\And
	{Stephan Stricker} \\
	PAIR Finance GmbH\\
	Berlin, Germany\\
	\texttt{stephan.stricker@pairfinance.com} \\
}

\date{June 3, 2021}

\renewcommand{\undertitle}{}

\begin{document}
\maketitle

\begin{abstract}
Based on debt collection agency (PAIR Finance) data, we developed a novel debtor typology framework by expanding previous approaches to 4 behavioral dimensions. The 4 dimensions we identified were willingness to pay, ability to pay, financial organization, and rational behavior. Using these dimensions, debtors could be classified into 16 different typologies. We identified 5 main typologies, which account for 63\% of the debtors in our data set. Further, we observed that each debtor typology reacted differently to the content and timing of reminder messages, allowing us to define an optimal debt collection strategy for each typology. For example, sending a reciprocity message at 8 p.m. in the evening is the most successful strategy to get a reaction from a debtor who is willing to pay their debt, able to pay their debt, chaotic in terms of their financial organization, and emotional when communicating and handling their finances. In sum, our findings suggest that each debtor type should be approached in a personalized way using different tonalities and timing schedules. 
\end{abstract}


\section{Introduction}
We are currently experiencing an increasing tendency to purchase products online rather than in stores. The global pandemic of COVID-19 has contributed to this trend, as many stores must close during lockdown periods. The increase of online purchases affects the payment behavior of customers, as they can delay payments through “buy now, pay later” options (e.g., paying via invoices). If the invoice is never paid or the direct debit bounces, companies face monetary losses due to defaulted payments. According to the Bundesverband Deutscher Inkasso-Unternehmen (BDIU), 20 million open claims are transferred to German debt collection agencies per year, and on average they recover 6 billion Euros per year \citep{bundesverband}. This does not only affect E-commerce companies, but also other industries such as insurance, mobility, financial services, shared economy, and telecommunication. 

Even though the economic importance is apparent, little is known regarding the underlying personality types of debtors or the most successful course of action to ensure timely repayment. Initial studies have shown that investigating debtors’ attitudes is important to have a broader understanding of the decision situation, which is crucial to avoid one-size-fits-all solutions \citep{dominykempson03}, \citep{watsonbarnao09}.
 
The current study makes two contributions to the literature. First, it used an extensive data set provided by a debt collection agency based in Germany; the data set includes a wide range of variables associated with a debtor’s financial situation, communication with the debt collection agency, and details about the open claim. This rich set of features allowed us to construct a debtor typology framework, and therefore provide a comprehensive overview of the possible drivers of defaulted payments. To our knowledge, the current study is the first that leveraged data collected by a debt collection agency to distinguish between different types of debtors. The second contribution the study makes is to evaluate the effectiveness of different communication strategies on the 16 types of debtors we defined. Specifically, we investigated how reactions to different combinations of message content and time of day differ across debtor types.

\section{Theoretical Background}
\label{sec:theoretical_background}

\subsection{Debtor personalities}

One of the first studies to closely investigate behavior of debtors was conducted by  \cite{livingstonelunt91}. They predicted personal debt using a range of variables, including demographic information, economic context, and psychological factors. Attitudinal factors towards money (e.g., considering credit as problematic) and psychological factors (e.g., locus of control, coping strategies, and consumer pleasure) were found to be important predictors of debt repayment. In the context of consumer debt, \cite{leawebleywalker95} found that it is debtors’ economic and demographic factors, and not their attitudinal or locus of control beliefs, that influenced debt repayment. Later, the first systematic debtor typology was developed, which differentiated debtor types depending on their ability to pay and their commitment to pay \citep{dominykempson03}. Based on these two dimensions, they identified four debtor types: 1) People withholding money on principle, 2) Ex-partners withholding payment, 3) People ‘working the system’, and 4) People ‘ducking responsibility’. Other studies have used the same two dimensions to classify types of debtors (e.g., \cite{gerardiherkenhoffohanianwillen18}). Similarly, in the context of student loan repayment, \cite{watsonbarnao09} identified four distinct groups of debtors based on attitudinal and behavioral data: Life indebted, Traditionalists, Entrepreneurs and Expedient Payees. 

In order to develop a more granular understanding of a debtor typology, the current study adapts a framework that is similar to one of the most popular measures of personality, namely the Myers-Briggs Type Indicator (MBTI: \cite{myers62}). It is based on the theory of psychological types introduced by Jung and differentiates 16 unique personality types. The test has been applied to a number of practical settings such as academic advising and career counseling (e.g., \cite{provostanchors87}), organizational behavior (e.g., \cite{bridges92}), and leadership (e.g., \cite{mccaulley90}). While the MBTI-framework has received optimistic \citep{mccaulley00} as well as skeptical reviews \citep{pittenger05}, a four-dimensional model is a promising approach to classify debtors into more granular personality types.

\subsection{Nudging financial decision making}

Several studies have investigated how financial decision making can be influenced (or nudged) in the context of tax repayment \citep{johnblume18}, saving strategy \citep{duretal21}, and consumer behavior \citep{demarqueetal15}. The concept and theory of nudging is based on the work of \cite{thalersunstein09} who proposed a list of nudges, aiming to adapt the choice architecture of a decision in a way that alters behavior in a predictable way. 

While there are many different nudges mentioned in the literature, one of the nudges most frequently used is based on providing information about the behavior of similar persons in the same situation, often referred to as social information. The concept is theoretically based on the Social Comparison Theory \citep{festinger54}. According to this theory, individuals have a fundamental motivation to evaluate their opinions and abilities by comparing themselves to similar others. If a discrepancy is detected, individuals experience a social pressure towards uniformity and an action will be taken to reduce the discrepancy. Previous findings showed that social comparison has been implemented as a mechanism to influence behavior in a wide range of different contexts, such as medical decision-making (e.g., \cite{hallsworthetal16}) and financial decisions (e.g., \cite{shangcroson09}). 

Furthermore, previous literature has identified reciprocity to be an important driver of decision-making. The concept of reciprocal altruism has its theoretical foundations in evolutionary psychology \citep{trivers71}, and it describes an individual's tendency to behave more cooperatively than predicted by the self-interest model in response to friendly actions by others \citep{fehrgaechter00}. \cite{ortegasanguinetti13} employed the norm of reciprocity to increase tax compliance in Venezuela.  

Beyond the informational content that is shared with the decision maker, another factor that has been found to have an influence on decision-making behavior is timing \citep{gillitzersinning20}. Depending on whether a person is faced with the decision situation in the morning or in the evening, the final choice can be different. For example, \cite{gulloetal19} studied the effect that the time of day has on consumer choices, and they observed that the basket of goods that shoppers buy tends to be less varied in the morning than at other times of day. In the context of judicial decisions, it was found that there is an increasing tendency to decide in favor of the status quo throughout the day \citep{danzigerlevavavnaimpesso11}. Regarding medical decisions, \cite{linderetal14} found that doctors tend to prescribe more antibiotics as the time of day increases. These findings indicate that the time of day could be a relevant factor when analyzing financial decision-making. On this basis, we took the time of day into consideration when investigating the effectiveness of different communication strategies.

\section{The study}

First, the goal of the current study is to establish a more granular framework of debtor typologies that extends the model of ability and willingness to pay by including psychological and organizational factors. The extensive data set provided by the debt collection agency allowed us to differentiate debtors from each other using a wide range of features. Second, we aim to investigate how communication strategies influence debtors’ decision-making. We are focusing on two dimensions of communication here, tonality and timing. Tonality refers to the content of the message sent to the debtors and timing refers to the time of day when an email was sent out. Specifically, we are interested in understanding how the success of these communication strategies differs depending on the debtors’ typology. The two success metrics we defined are reactions to outbound messages and any payment to settle the open debt.

\section{Methodology}

\subsection{Subject pool}

The open claims are associated with purchases at different companies in a wide range of industries, including E-commerce, financial services, insurance services, mobility, shared economy, and telecommunication. The debtor typology framework established here is based on the data of 391,266 debtors and 1,673,759 actions of outbound communication. The debtors selected for this study started the debt collection process between January 1st, 2017 and September 30th, 2020.The most frequently represented clients are E-commerce companies based in Germany. In this debtor pool, 57\% are female, 34\% are male, and 9\% are unknown. Age information is available for 61\% of the debtors and the average age is 35 years (\textit{SD} = 12.3). The debtors analyzed in this study are exclusively German. 

\subsection{Typology framework}

The process of establishing our framework for debtor typologies can be broken down into  three phases:

\subsubsection{Identification of dimensions}

First, we identified dimensions across which debtors could differ (e.g., “Willingness to pay”). The dimensions were established based on existing literature on different typologies in financial behaviour, as well as by conducting interviews with debt collection agents to understand how they experience differences between debtors based on their interactions and behaviour. Furthermore, we analyzed available data in the debtor pool that was available to us to understand the data points along which debtors can be differentiated.
Based on our analysis, four dimensions were identified: willingness to pay (willing vs. defiant), ability to pay (able vs. insolvent), financial organization (chaotic vs. organized), and rational behavior (rational vs. emotional). Rational behavior is defined in terms of rational financial decision making (avoiding additional fees) and rational communication (writing structured emails, avoiding emotional language). The first two dimensions are commonly used in literature on debt collection (\cite{dominykempson03}; \cite{gerardiherkenhoffohanianwillen18}), while the latter two are, to our knowledge, herein analyzed in the context of debtors for the first time.

\subsubsection{Identification of features for each dimension}

Second, for each of the four dimensions, we generated a list of features and data points that were relevant to compile a score on this dimension for each debtor. The features were identified based on a combination of previous findings in the literature and the internal debt collection expertise of the company. Members of the collection team were interviewed and their extensive experience in direct communication with debtors was used to identify the respective features. Overall, we identified 21 relevant features to estimate willingness to pay, 19 features to estimate payment ability, 15 features to estimate financial organization, and 17 features to estimate rational behavior. Out of the 72 identified features, 19 features were available \textit{ex ante} (upon starting the collection process and before any communication was sent out or received). A full list of the dimensions and features can be found in the appendix (Tables~\ref{tab:willingness},~\ref{tab:ability},~\ref{tab:organization},~\ref{tab:rationality}). 

\subsubsection{Generation of typology scores}

Third, based on the compiled features, we generated scores for each debtor on each dimension. Several factors were taken into account to determine the weight assigned to each feature. For one, certain features were empirically determined to be more important indicators of a dimension than others. Additionally, we reduced weights for features that would only affect a small fraction of debtors while increasing weights for features with a strong differentiative power. This way, we prevented heavily skewed score distributions or the presence of strong modes in the scores. Some features are binary (e.g., “Is the email address valid?”) and others are continuous (e.g., “How many replies did we receive?”). Formally, the scores were generated by:

$$ y_d = \sum_{i=1}^{N_d} w_{i,d} x_{i,d}$$

where $d$ is one of the four dimensions, $w_{i,d}$ the weight of feature $x_{i,d}$ for dimension $d$ where $1 \leq i \leq N_d$. All features $x_{i,d}$ are assumed to be standardized prior to the scoring across dimension $d$. Eventually, we normalized all scores as following, which ensured a distribution between $0$ and $1$ across our debtor pool:

$$\widetilde{y_d} = \frac{y_d - min\{y_d\}}{max\{y_d\} - min\{y_d\}}$$

In the end, we characterized each debtor by four scores, where values ranged between 0 and 1 (inclusive). Next, we used the typology scores to distinguish between different types of debtors. To do this, we assigned debtors to a specific type using 0.5 as the cut-off in the continuous range of scores described above. Therefore, debtors with scores equal to or above 0.5 on the willingness to pay scale were classified as “willing”, while those with scores below 0.5 were classified as “defiant”. On the ability to pay scale, debtors with scores equal to or above 0.5 were classified as “able”, while those with scores below 0.5 were classified as “insolvent”. We followed the same procedure for the scales of financial organization and rational behavior.

Following this procedure, we labelled each debtor so that it would belong to one of the 16 resulting typologies. For instance, the WACE typology is defined by people who are Willing, Able, Chaotic and Emotional (see Table~\ref{tab:dimensions}).

\begin{table}[ht]
	\caption{Overview of the 16 debtor typologies. The 16 types are generated using a combination of four dimensions. Each letter represents the classification for each of the four dimensions.}
	\centering
	\begin{tabular}{cccccccc}
		\toprule
		\multicolumn{8}{c}{Payment ability}  \\
		\midrule
		& & Able & Able & Insolvent & Insolvent & & \\
		\cmidrule{2-7}
		\multirow{4}{*}{\shortstack[c]{Willingness \\ to pay}} & Willing & WAOR & WACR & WICR & WIOR & Rational & \multirow{4}{*}{\shortstack[c]{Rational \\ behavior}} \\
		\cmidrule{3-6}
		& Willing & WAOE & WACE & WICE & WIOE & Emotional & \\
		\cmidrule{3-6}
		& Defiant & DAOE & DACE & DICE & DIOE & Emotional & \\
		\cmidrule{3-6}
		& Defiant & DAOR & DACR & DICR & DIOR & Rational & \\
		\cmidrule{2-7}
		& & Organized & Chaotic & Chaotic & Organized & & \\
		\midrule
		\multicolumn{8}{c}{Financial organization}  \\
		\bottomrule
	\end{tabular}
	\label{tab:dimensions}
\end{table}

\subsection{Procedure}

As part of the company’s debt collection process, we were able to collect 72 data points to better understand the differences between debtor types. At the beginning of the collection process, debtors were contacted via email or letter to remind them about the outstanding debt. Messages varied in the way they were phrased (tonality) and the time of day when they were sent out (timing). Note that timing only relates to sending emails, since the timing of letters cannot be influenced. While the company uses a number of channels to contact debtors (e.g., SMS and WhatsApp messages), our analysis focused on emails and letters. The tonalities analyzed in this study include: a friendly and understanding message (cooperative); a message with neutral tone only including the most relevant information (informative); a message which highlights negative consequences and thus puts more pressure on debtors (hard); a message which informs about the relevant behavior of a social reference group (social comparison); and finally, a message that highlights the effort employees of the company where the debt is owed have put in to offer the respective product (reciprocity). Examples for each tonality can be found in the appendix (Table~\ref{tab:templates_translated}). Messages were sent out either early in the morning (between 6 a.m. and 10 a.m.), around noon (between 10 a.m. and 2 p.m.), in the afternoon (between 2 p.m. and 6 p.m.), or in the evening (between 6 p.m. and 10 p.m.). These time slots are further referred to as “8:00”, “12:00”, “16:00”, and “20:00”.

In our data, tonalities and timing of the outbound communication were assigned to debtors predominantly based on the recommendations of a machine learning algorithm and in few cases based on business requirements. The machine learning procedure that was used is known as contextual bandits, which is a reinforcement learning approach. This approach works according to a trade-off between exploration and exploitation. It explores to try out how well debtors react when receiving a certain kind of message and exploits by preferably serving the action expected to be optimal. Given the exploration method of the algorithm, we have the guarantee that all different options for tonality and timing are regularly being tried out. After messages were sent, we recorded the reactions taken by debtors including visits to the payment page, inbound communication, requests for payment solutions, partial and full payments. Reactions were assigned to the messages that were sent most recently to the debtor.

\subsection{Data Analysis}

First, we analyzed the distribution of typologies in the subject pool. To do this, we used descriptive statistics and computed the relative frequency of each typology on a debtor level. Second, to investigate the relationship between typology and both the tonality and the timing of the message, we analyzed debtors’ reaction and payment rates. Reactions include any payment page visit, taking an instalment plan or making a promise to pay, any inbound communication or any direct payment (partial or full). For the analyses including tonality and timing, we had to exclude any data collected prior to May 1st, 2020, given that we only introduced the reciprocity tonality at that time. Debtors where none of the five tonalities described above were used also had to be discarded. The distribution of the tonalities in our data set can be found in the appendix (Figure~\ref{fig:distribution_tonalities}). Finally, we also limited the analysis to the first stages of the debt collection process, before any legal steps were taken in case of non-payment. This process takes 6-8 weeks. This exclusion procedure left us with a final sample of 80,706 debtors and 362,462 actions of outbound communication for the analyses regarding tonality and timing. The exclusion procedure did not affect the distribution of age and gender in the subject pool (58\% females, \textit{M} = 35 years, \textit{SD} = 12.6). 

To assess the success of a communication strategy, we primarily focus on the reaction rate of the debtors. We believe it is a highly relevant metric for debt collection since recovery of the open claim usually is a lengthy process where debtors engage in a funnel that eventually leads to the full payment. In this funnel, getting a debtor to react is one of the most critical steps and especially important when it comes to the influence of message content and timing. We consider reactions the most direct type of information about the effect of a message, whereas payments can potentially be confounded with other factors such as the user interface of the payment page. Thus, we focus on reporting results for reactions but nonetheless the corresponding analyses using payment as a dependent variable can be found in the appendix (Figures~\ref{fig:payment_tonality},~\ref{fig:payment_timing}).

\section{Results}

\subsection{Distribution of the typologies}

\begin{figure}[ht]
	\centering
	\caption{Distribution of debtor typologies displaying the frequency each typology occurs in the subject pool in percent.}
	\includegraphics[width=15cm, height=10cm]{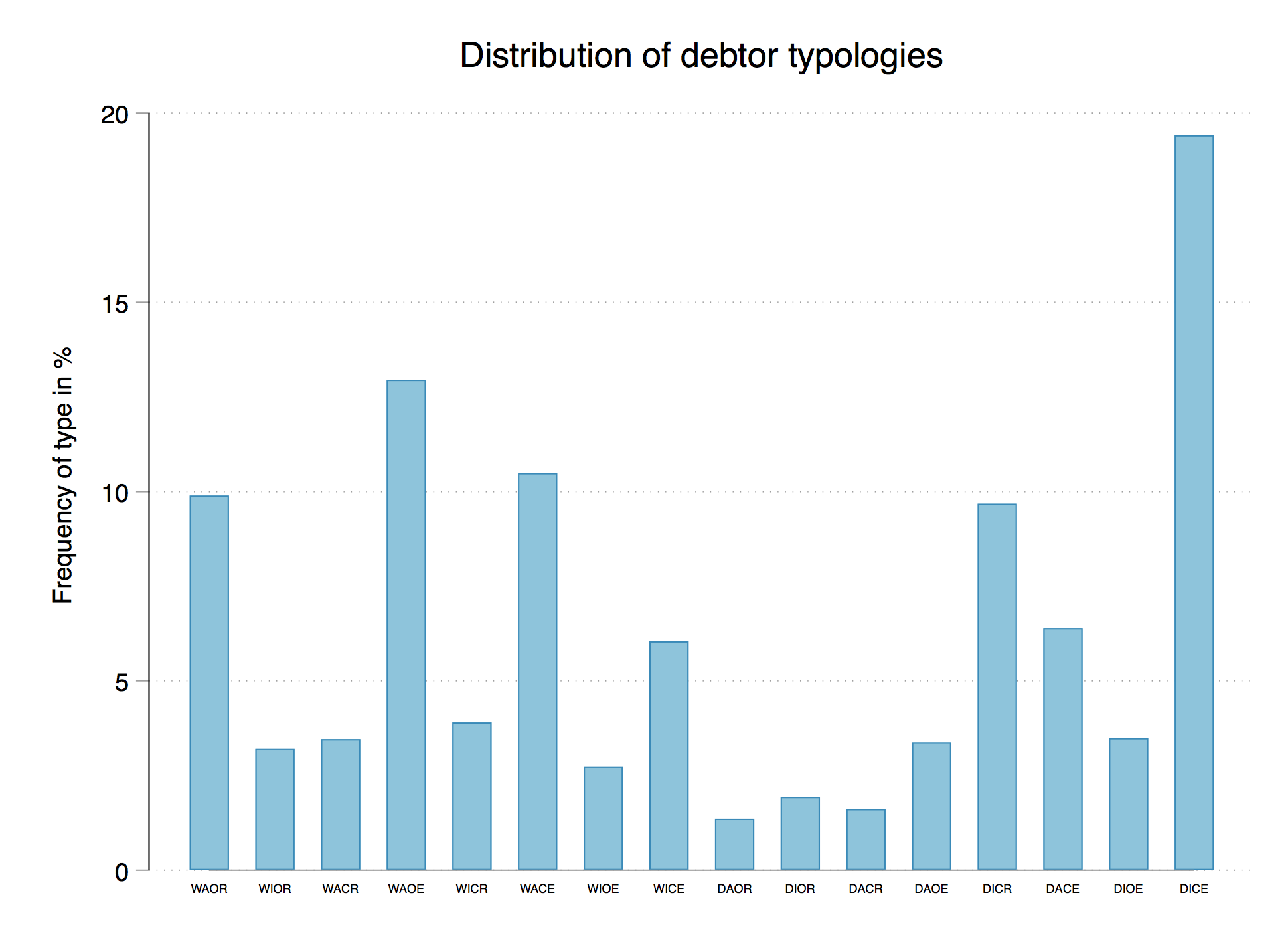}
	\label{fig:distribution_typologies}
\end{figure}

First, we analyzed the distribution of debtor types in our subject pool (see Figure~\ref{fig:distribution_typologies}). From this chart, we can identify five main typologies which account for about 63\% of the debtors in our data.

\begin{itemize}
	\item DICE: Defiant-Insolvent-Chaotic-Emotional. Typically, these are debtors who tend to refuse any payment solution and are in a difficult financial situation. Many of them have already been debtors in the past and they do not engage in a conversation with the debt collection agency, despite additional incentives to trigger reactions.
	\item WAOE: Willing-Able-Organized-Emotional. Typically, these are debtors who tend to opt for a payment solution and who are in a better financial situation, who quickly react to messages from debt collection even though their communication can sometimes be erratic.
	\item WACE: Willing-Able-Chaotic-Emotional. Typically, these are debtors who tend to opt for a payment solution and who are in a better financial situation. However, they are not able to stick to payment schedules and are not factual in their communication.
	\item WAOR: Willing-Able-Organized-Rational. Typically, these are debtors who tend to opt for a payment solution and who are in a better financial situation, who display a more pronounced digital behavior, and who make rational financial decisions by paying in reaction to reduction offers.
	\item DICR: Defiant-Insolvent-Chaotic-Rational. Typically, these are debtors who tend to refuse any payment solution and are in a difficult financial situation, who do not engage in a conversation with the debt collection agency and who entered debt collection with large open claims.
\end{itemize}

These five main typologies highlight the fact that willingness and ability to pay are strongly correlated. On the other hand, the rationality score is the score that is most decorrelated from the three other dimensions.

\subsection{How do different typologies react?}

Based on the time frame considered in this study, 60\% of the debtors reacted to our messages (the reaction rate for the full collection process reaches 77\%). Breaking the reaction rates further down for the debtor typologies, the results showed that some types were more likely to react to reminder messages than others. For example, the WAOR typology is the typology that reacts most frequently (98\%) whereas the DICE typology is the least reactive (20\%). This largely comes as a consequence of the definition of these typologies. Table~\ref{tab:overall_reaction_rates} shows the reaction rates for each of the typologies (the corresponding table for payments can be found in the appendix, Table~\ref{tab:overall_payment_rates}). Interestingly, reaction rates are considerably higher for willing debtors than for defiant debtors, with the exception of the DAOR debtors who are also very reactive (80.2\%).   

\begin{table}[ht]
	\caption{Overview of reaction rates by debtor to outbound messages for each typology}
	\centering
	\begin{tabular}{lr}
		\toprule
		Typology     & Reaction rate (in \%)  \\
		\midrule
		DACE & 45.1 \\
		DACR & 58.4 \\
		DAOE & 61.5 \\
		DAOR & 80.2 \\
		DICE & 19.8 \\
        DICR & 27.6 \\
        DIOE & 32.4 \\ 
        DIOR & 51.7 \\
        WACE & 83.0 \\
        WACR & 93.7 \\
        WAOE & 90.1 \\
        WAOR & 97.8 \\
        WICE & 78.8 \\
        WICR & 80.1 \\
        WIOE & 84.5 \\
        WIOR & 91.9 \\
		\bottomrule
	\end{tabular}
	\label{tab:overall_reaction_rates}
\end{table}

\subsection{How do different typologies react to different message tonalities?}

\subsubsection{Tonality}

Taking the message content into account, we further observed that the tonality of the reminder messages also influenced reactions heterogeneously across typologies. For each typology, we analyzed the proportion of reactions to outbound messages depending on the tonality that was used. A reaction is always assigned to the tonality of the most recently sent message. While the choice of tonalities in our sample was not randomized, the fact that we are comparing results by typology and therefore controlling for differences in debtor characteristics (a procedure sometimes called blocking) allows us to draw meaningful conclusions. This methodological aspect is discussed further in the discussion section.

\begin{figure}[ht]
	\centering
	\caption{Display of reaction rates for each tonality depending on typology.}
	\includegraphics[width=18cm, height=6cm]{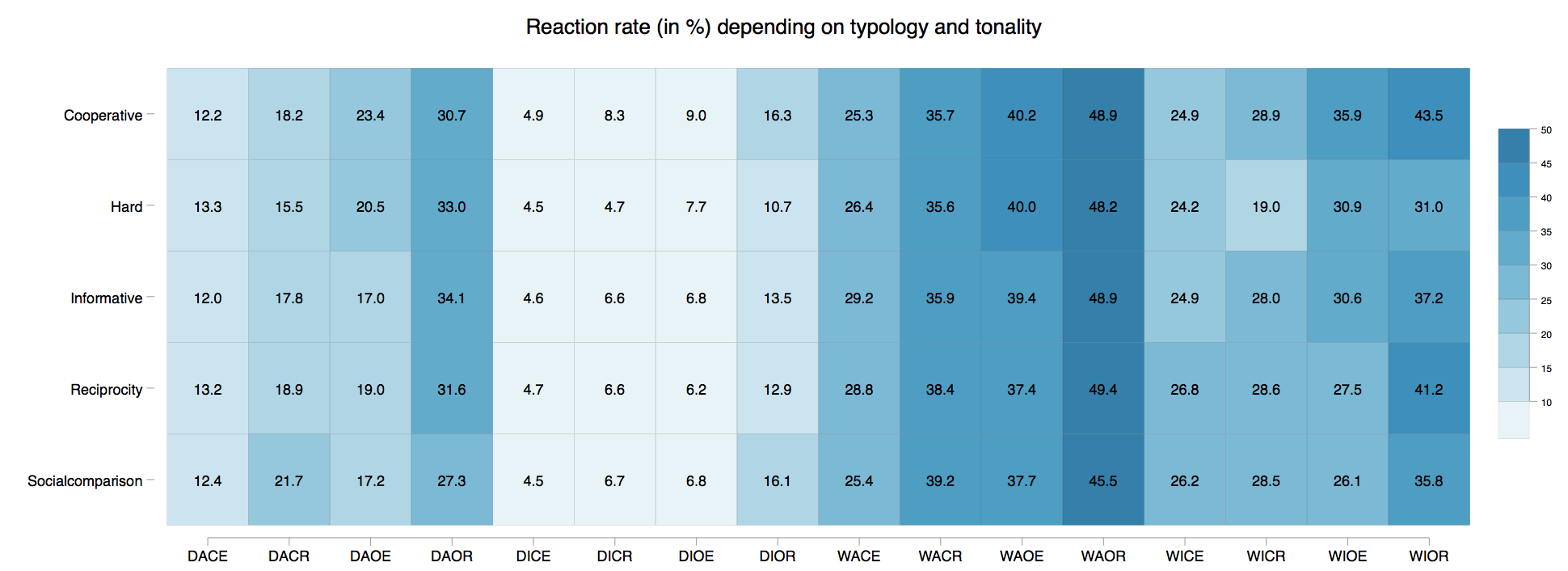}
	\label{fig:reaction_tonality}
\end{figure}

Figure~\ref{fig:reaction_tonality} represents the reaction rates by email tonality and by debtor typology (the equivalent table for payment rates can be found in the appendix in Figure~\ref{fig:payment_tonality}). The reaction rates shown in this figure represent the percentage of debtors of a given typology who reacted after a message of a given tonality. For instance, WAOR debtors react in 49.4\% of the cases after they received a reciprocity email. However, they react in only 45.5\% of the cases after they received a social comparison email. 

\begin{itemize}
    \item DICE: This group of debtors is the least likely to react to outbound messages. The cooperative tonality is the most successful in achieving reactions whereas the hard tonality is the least successful. 
    \item WAOE: We can see that two tonalities are standing out as being equally successful: cooperative and hard. The reciprocity tonality reduces reaction rates by 3\%.
    \item WACE: The two best tonalities are the informative and the reciprocity tonality, which achieve the highest reaction rates. On the other side of the spectrum, we see that cooperative is performing less well, with reaction rates dropping by 4\%.
    \item WAOR: The reciprocity tonality appears to be the most effective, closely followed by informative and cooperative. Social comparison, on the other hand, underperforms with reaction rates 4\% lower.
    \item DICR: The cooperative tonality seems to be working best for these debtors, whereas hard tonality reduces reaction rates by about 4\%. This suggests that for debtors who are in difficulty, the hard tonality proves to be counterproductive, and that it is beneficial to be more cooperative. This is a result that we can also clearly observe for the other defiant and insolvent typologies (DICE, DIOE and DIOR).
\end{itemize}

The results of the chi-squared test indicate that we can reject the null hypothesis stating that the number of reactions across different tonalities and the typology are independent, $\chi^2 (60, N = 64,557) = 451.05, p < .001$. Results for payments can be found in the appendix (Table~\ref{tab:chi2payments_tonality}). 
These results indicate that it is crucial to tailor the communication strategy to each specific kind of debtor. By serving standard or default communications to all the debtors, it appears that debt collection agencies are bound to opt for a sub-optimal strategy, failing to nudge as many people as possible to reimburse their debts. Our results also help to debunk some of the preconceived ideas about debt collection management, that debtors who are unlikely to react should receive hard, demanding, and even threatening messaging; according to the data, such debtors can actually benefit from a more accommodating tone, because adopting a hard and threatening tone will prove to further deter them. This highlights the tremendous potential of using data to profile debtors and to determine the best way to communicate with them in order to get a reaction first, and later down the funnel a full payment. 

\subsubsection{Tonality and Timing}

\begin{figure}[ht]
	\centering
	\caption{Display of reaction rates for each time-tonality combination depending on typology. }
	\includegraphics[width=18cm, height=9cm]{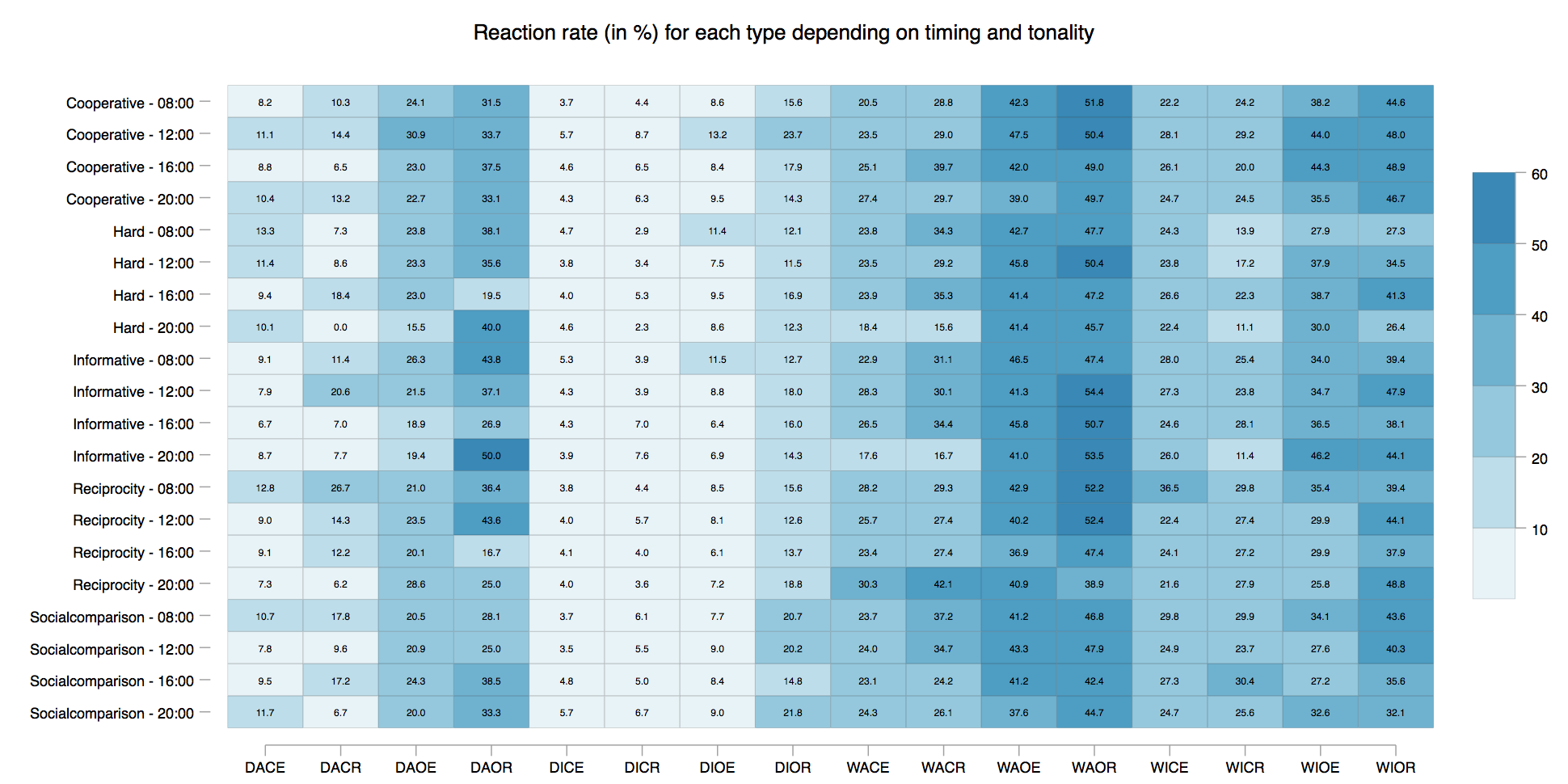}
	\label{fig:reaction_timing}
\end{figure}

This analysis can be pushed even further by considering at what time, in combination with which tonality, the debtors should be approached to achieve more reactions and payments. To analyze this, we considered four times in the day: 8:00, 12:00, 16:00, and 20:00 (for this analysis we only use email communication, since we cannot control the time of the day when a debtor receives the letters). A payment or a reaction is always assigned to the time-tonality combination of the most recently sent message. Figure~\ref{fig:reaction_timing} represents the reaction rates by time-tonality combination and by debtor typology (the equivalent figure for payment rates can be found in the appendix in Figure~\ref{fig:payment_timing}). The reaction rates shown in this figure represent the percentage of debtors of a given typology who reacted after a message of a given tonality at a given time. The results reveal that there are differences in how successful a specific tonality is in terms of subsequent reactions, depending on the time of day. For example, the DICE and WAOE debtors both react most frequently when a cooperative message is sent to them at 12 p.m. (5.7\% and 47.5\% respectively). For WACE debtors, the results show that in combination with sending a message at 8 p.m. in the evening, a reciprocity message is associated with the highest reaction rate (30.3\%). The WAOR debtors react in 54.4\% of the cases after receiving an informative message at around 12 p.m., while they only react in 47.4\% of the cases when receiving the same message in the morning. Finally, for DICR debtors, we observe reactions in 8.7\% of the cases after sending a cooperative message around 12 p.m., while the reaction rate for the same message drops to 4.4\% around 8 a.m.

Again, the chi-squared test enables us to reject the null hypothesis that the typology and the timing-tonality combinations are independent in terms of observed reaction counts, $\chi^2 (285, N = 42,843) = 745.19, p < .001$. Find the respective results for payments in the appendix (Table~\ref{tab:chi2payments_timing}).

\section{Discussion}

First, the study shows that using such an extensive data set offers the possibility to distinguish 16 different debtor typologies based on willingness to pay, ability to pay, financial organization, and rational behavior. The most represented typology we observed in our data set were debtors who are defiant to pay their debt, unable to pay, chaotic in terms of their financial organization and emotional when communicating and handling their financials (DICE). Other frequent typologies included debtors who are willing to pay, able to pay, organized, and emotional (WAOE); willing to pay, able to pay, chaotic, and emotional (WACE); willing to pay, able to pay, organized, and rational (WAOR); defiant to pay, insolvent, chaotic, and rational (DICR). Second, our results indicate that depending on the debtor typology a different outbound communication strategy is most successful in terms of leading to reactions and payments. Specifically, for the most frequent debtor typology (DICE), we found that the most successful strategy is sending a cooperative message at 12 p.m. in the afternoon. 

It is important to note that there are three methodological challenges associated with this study, which are the natural consequences of the applied nature of this research project. The first limitation is the nature of the debtor pool we analyzed. We know, for instance, that the debtor pool primarily represents digital consumers, which might have in turn skewed the demographics in the study (age, gender, level of income, etc.). Overall, the results of this study should be applied to subject groups with similar demographics. 

The second limitation comes from the way we computed the four typology scores for each debtor. To do this, we were constrained by the data available to us. For instance, we tried to define the ability to pay without having any information about a debtor’s income or the exact state of their liquidity. Instead, we had to build proxies with the data we had using some assumptions. Once we defined a list of relevant features for each typology dimension, we combined and weighed them to compute the typology scores. These weights were set empirically as described in the methodology section, and while some decisions with respect to these weights were determined \textit{post hoc}, we believe that our methodology fulfills its purpose– to define two opposite groups along each dimension that exhibit different behaviors.

The third limitation affects the analysis of the interactions between typologies and tonalities. It can be objected that tonalities were not assigned in a randomized fashion, making the interpretation of our results difficult. This is a strong limitation to our work given that the analysis we conducted was \textit{ex post} in an applied setting. However, we believe we have taken some steps that helped minimize the impact of the lack of randomization. First, we reduced the time frame of the analysis for tonality and timing between May and September 2020, where each of the five tonalities were assigned to outbound communications. Second, we excluded any legal communication points, which typically only concern reluctant debtors who still refuse to pay, to eliminate biases due to the use of certain tonalities in stages where reaction rates are naturally much lower. Third, the reaction rates we analyzed are within each typology, which highly reduces any underlying biases by comparing similar debtors to each other. This technique has often been recommended in the literature to analyze the outcome of non-randomized experiments and is often known as \textit{blocking} or \textit{subclassification}.

We encourage future studies to remove some of the limitations mentioned above. It would be interesting to enrich the definition of the typology dimensions with more data sources that relate more directly to willingness, ability, organization and rationality, such as financial data or personality surveys. Further steps should also involve validating these results with a randomized experiment, which would involve setting up an experiment infrastructure \textit{ex ante} to remove potential hidden biases. Future studies should also investigate other debt collection strategies, for instance if different typologies react in the same way to financial incentives or financial solutions. We believe there is a wide range of possibilities that remains to be explored in this area.

\section{Conclusion}

In this study, we proposed a new framework to analyze the behaviour of debtors based on typologies. Based on the four dimensions -ability to pay, willingness to pay, organization, and rationality- we defined 16 debtor typologies. While the literature has frequently focused on the first two dimensions, we believe that better insights and business results can be achieved by developing even more dimensions. The main benefit is that such classification schemes enable companies dealing with receivable management to better understand debtors and tailor their approach to each debtor profile in a more granular way. Our results suggest that each debtor type should be approached in a personalized way using different tonalities and timing schedules. This granular strategy is a shift from the monolithic standard approach commonly used to collect debt, leveraging data to achieve better recovery results.

\bibliographystyle{plainnat}


\newpage

\section*{Appendix}

\setcounter{table}{0}
\renewcommand{\thetable}{A\arabic{table}}

\begin{table}[ht]
	\caption{List of features and associated weights for willingness to pay}
	\centering
	\begin{tabular}{lr}
		\toprule
		Feature     & Weight  \\
		\midrule
		Did the debtor ask for an instalment plan via email? & 2   \\
		Did the debtor ask for a payment pause via email?     & 2     \\
		Did the debtor commit to paying the debt on a specific date at any time? & 0.3 \\
		Did the debtor request an instalment plan (without signing it yet) at any time? & 0.8 \\
		Was any payment solution taken (e.g., instalment plan, payment pause)? & 0.7 \\
		Did the debtor request and sign an instalment plan at any time? & 1.6 \\
		Did the debtor make a partial payment within 60 days? & 1.7 \\
		Was the payment page visited? & 0.6 \\
		Is there debt counseling involved in the process? & 1 \\
		Is the name of the debtor represented in the email address? & 0.5 \\
		Is it a fraudulent case file? & -0.5 \\
		How old is the debt? & -1.5 \\
		Do we observe any reaction from the debtor? & 0.3 \\
		How long did it take until we observed the first reaction? & -1.7 \\
		Was the debt disputed? & -1.2 \\
		Is the debtor’s name valid? & 0.6 \\
		Is the debtor’s email address valid? & 0.8 \\
		Is the debtor’s address valid? & 0.6 \\
		Was a court process initiated for this case file? & -0.5 \\
		Were multiple solutions chosen in the process (signaling less willingness)? & -3.5 \\
		Did the debtor choose to pay via direct debit? & 0.5 \\
		\bottomrule
	\end{tabular}
	\label{tab:willingness}
\end{table}

\begin{table}[ht]
	\caption{List of features and associated weights for ability to pay}
	\centering
	\begin{tabular}{lr}
		\toprule
		Feature     & Weight  \\
		\midrule
        What is the PAIR Score (internal scoring system)? & 1.3 \\
        Was the PAIR Score high? & 1.5 \\
        How large is the open claim with regards to the debtor’s rent price? & -1 \\
        What is the SCHUFA score? & 2 \\
        Was the SCHUFA score high? & 2.2 \\
        Was the debt paid? & 1.5 \\
        Was any payment solution taken? & -1.3 \\
        Has the debtor died/is in prison? & -0.5 \\
        Did the debtor initiate an insolvency process? & -1 \\
        Was more than one technical device used? & 1.2 \\
        Was an OS / Mac device used? & 1.6 \\
        What is the rent price of the debtor’s residential area? & 0.5 \\
        What is the unemployment ratio of the debtor’s residential area? & -0.5 \\
        What is the disposable income of the debtor’s residential area? & 0.7 \\
        What is the size of the main claim? & -2.5 \\
        there debt counseling involved in the process? & -0.5 \\
        Is the debtor a recurrent debtor? & -1 \\
        Was a court process initiated for this case file? & -1 \\
		\bottomrule
	\end{tabular}
	\label{tab:ability}
\end{table}

\begin{table}[ht]
	\caption{List of features and associated weights for financial organization}
	\centering
	\begin{tabular}{lr}
		\toprule
		Feature     & Weight  \\
		\midrule
        Did the debtor stick to his/her commitment to pay the debt on a specific date? & 1 \\
        Did the debtor stick to the instalment plan schedule? & 1.3 \\
        Is the name of the debtor represented in the email address? & 0.5 \\
        Was there ever a late payment in an instalment plan? & 0.7 \\
        Was a previous instalment plan cancelled? & -0.5 \\
        Was there more than one payment solution (e.g. instalment plan) chosen? & -0.5 \\
        How long did it take the debtor to sign the instalment plan after agreeing on it? & -0.5 \\
        Is there debt counseling involved in the process? & -0.5 \\
        Did the debtor enter debt collection because the purchased item was returned too late? & -1 \\
        Did the debtor pay the debt to the creditor directly? & -0.7 \\
        Was there ever another debt collection case for this debtor? & -0.8 \\
        Has the payment attempt expired? & -0.5 \\
        Is the debtor’s email address valid? & 1 \\
        Did the debtor use attachments when sending emails? & 1 \\
        Did the debtor choose to pay via direct debit? & 0.5 \\
		\bottomrule
	\end{tabular}
	\label{tab:organization}
\end{table}

\begin{table}[ht]
	\caption{List of features and associated weights for rational behavior}
	\centering
	\begin{tabular}{lr}
		\toprule
		Feature     & Weight  \\
		\midrule
        Was there any insulting language used when the debtor sent an email? & -1 \\
        Was there any repeated punctuation (i.e. "!!!" or "???") used when the debtor sent an email? & -0.8 \\
        Was a formal greeting and ending included when the debtor sent an email? & 1.3 \\
        Were the emails very short (<20 words) or very long (>100 words)? & -1 \\
        Did the debtor use attachments when sending emails? & 1.5 \\
        Were there any emojis used when the debtor sent an email? & -0.5 \\
        Were multiple emails sent in response to one email? & -1.5 \\
        What is the proportion of uppercase characters in the emails sent by the debtor? & -1.5 \\
        Did the debtor pay after the debt collection fee was increased? & -2.3 \\
        What is the ratio of the debt amount to the debt collection fee? & 8 \\
        Did the debtor take a payment pause after the debt was reduced? & 2.7 \\
        Did the debtor take an instalment plan after the debt was reduced? & 2.6 \\
        Did the debtor pay in reaction to a reduction of the debt? & 2.5 \\
        Did the debtor take an instalment plan without paying additional fees? & 2.3 \\
        Did the debtor take a payment pause without paying additional fees? & 2 \\
        Did the debtor end up paying the debt in the court process? & -1 \\
		\bottomrule
	\end{tabular}
	\label{tab:rationality}
\end{table}

\begin{table}[ht]
	\caption{Translation of the templates for the different tonalities that were used in outbound messages.}
	\centering
	\begin{tabularx}{\textwidth}{ lX }
		\toprule
		Tonality     & Content  \\
		\midrule
        Cooperative & Maybe you simply did not see the invoice, or maybe you forgot to pay it due to a hectic everyday life situation. No matter what the reason is, surely this was not done on purpose. Thus, we now offer you the possibility to pay the outstanding amount of [claim amount] simply via one of the following payment solutions. Click on the respective symbol to select one of the solutions. \\
        \midrule
        Hard & Despite several reminders from [creditor], you still have not paid back your outstanding claim. We have now been mandated to manage this case, which results in additional costs. We urge you to pay the outstanding claim of [claim amount] without any further delay. Please use the following link to do so: \\
        \midrule
        Informative & We ask you to pay your open claim now and offer you the possibility to simply pay the outstanding amount of [claim amount] via the following link: \\
        \midrule
        Reciprocity & Behind [creditor], there is a driven team of people who work hard to offer you their services and products. You can surely understand that paying your bills is necessary to maintain the quality service you value as a customer.  \\
        & Therefore, we expect that you will fulfill your obligation towards [creditor] and pay the outstanding amount of [claim amount] due. To do so, you can simply use the following link: \\
        \midrule
        Social comparison & 9 out of 10 people pay their bills on time in Germany. As long as your claim of [claim amount] is still outstanding, you belong to a small group of customers who do not pay their bills. Change this situation and use the following link to make the payment: \\
		\bottomrule
	\end{tabularx}
	\label{tab:templates_translated}
\end{table}

\begin{table}[ht]
	\caption{Overview of payment rates by debtor to outbound messages for each typology}
	\centering
	\begin{tabular}{lr}
		\toprule
		Typology     & Payment rate (in \%)  \\
		\midrule
		DACE & 21.6 \\
		DACR & 25.9 \\
		DAOE & 22.6 \\
		DAOR & 39.2 \\
		DICE & 4.8 \\
        DICR & 5.3 \\
        DIOE & 4.9 \\ 
        DIOR & 8.4 \\
        WACE & 74.3 \\
        WACR & 87.5 \\
        WAOE & 79.6 \\
        WAOR & 94.8 \\
        WICE & 58.4 \\
        WICR & 59.2 \\
        WIOE & 62.7 \\
        WIOR & 75.5 \\
		\bottomrule
	\end{tabular}
	\label{tab:overall_payment_rates}
\end{table}

\begin{table}[ht]
	\caption{Chi-squared test for the number of payments in reaction to different tonalities depending on a debtor’s typology.}
	\centering
	\begin{tabular}{llll}
		\toprule
		Degrees of freedom & N & $\chi^2$ & p-value \\
		\midrule
        60 & 37,362 & 265.29 & < .001 \\
		\bottomrule
	\end{tabular}
	\label{tab:chi2payments_tonality}
\end{table}

\begin{table}[ht]
	\caption{Chi-squared test for the number of payments to different combinations of timing and tonalities depending on a debtor’s typology.}
	\centering
	\begin{tabular}{llll}
		\toprule
		Degrees of freedom & N & $\chi^2$ & p-value \\
		\midrule
        285 & 23,853 & 539.26 & < .001 \\
		\bottomrule
	\end{tabular}
	\label{tab:chi2payments_timing}
\end{table}

\setcounter{figure}{0}
\renewcommand{\thefigure}{A\arabic{figure}}

\begin{figure}[ht]
	\centering
	\caption{Distribution of the different tonalities used in outbound communication. The time frame is May 1st, 2020 until September 30th, 2020.}
	\includegraphics[width=16cm, height=11cm]{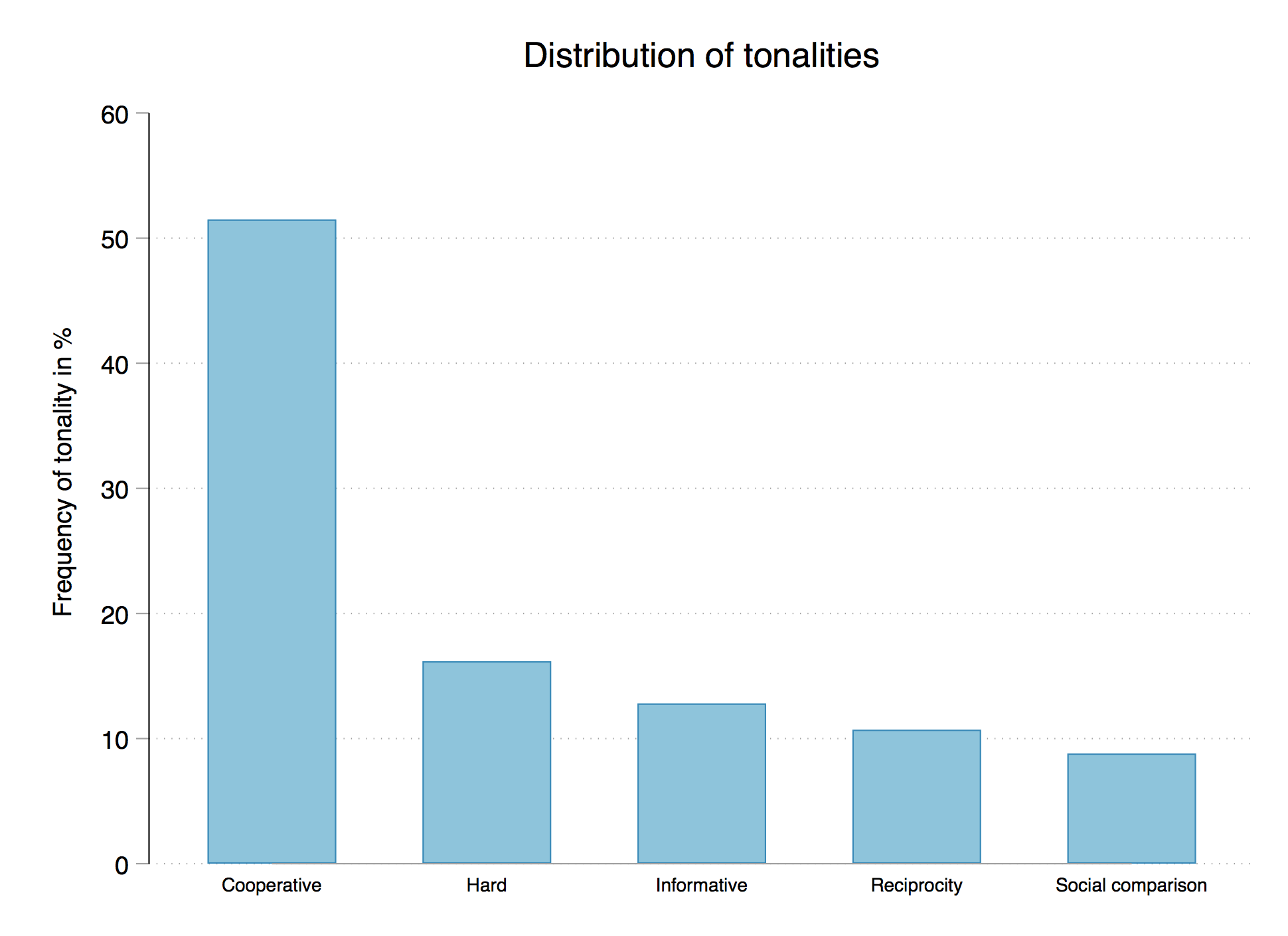}
	\label{fig:distribution_tonalities}
\end{figure}

\begin{figure}[ht]
	\centering
	\caption{Display of payment rates for each tonality depending on typology. Payments include partial as well as full payments.}
	\includegraphics[width=18cm, height=6cm]{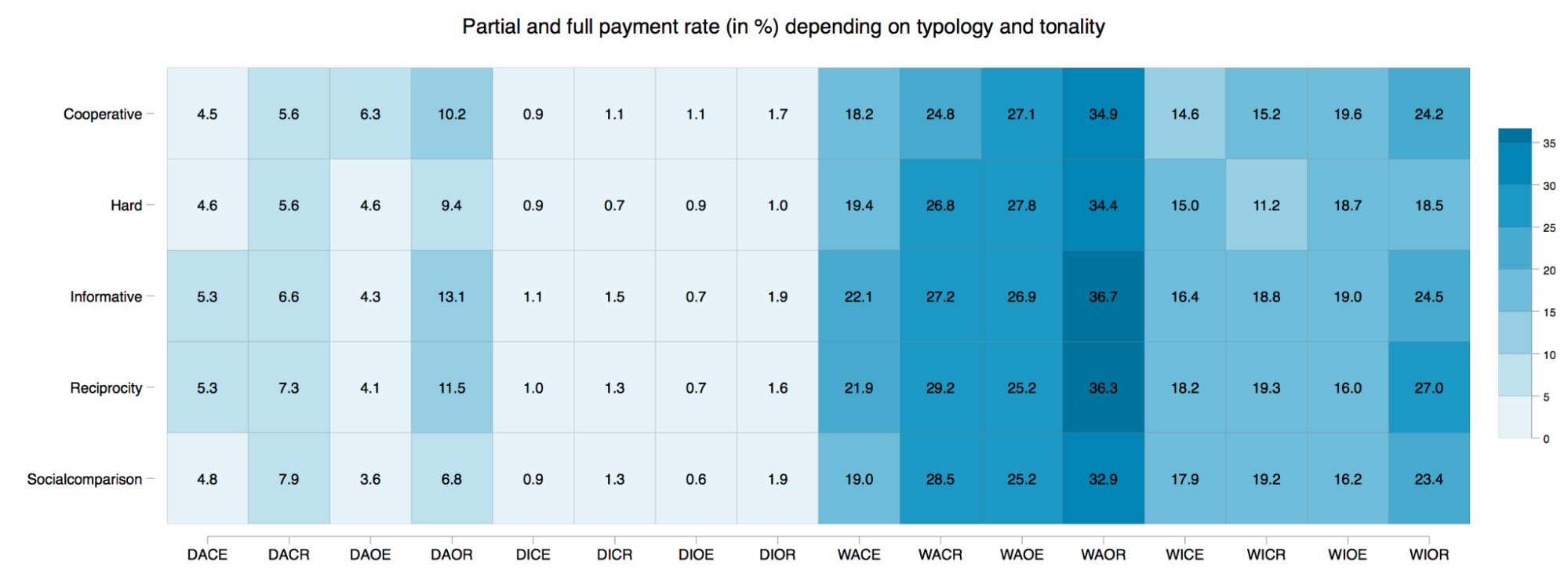}
	\label{fig:payment_tonality}
\end{figure}

\begin{figure}[ht]
	\centering
	\caption{Display of partial and full payment rates for each time-tonality combination.}
	\includegraphics[width=18cm, height=9cm]{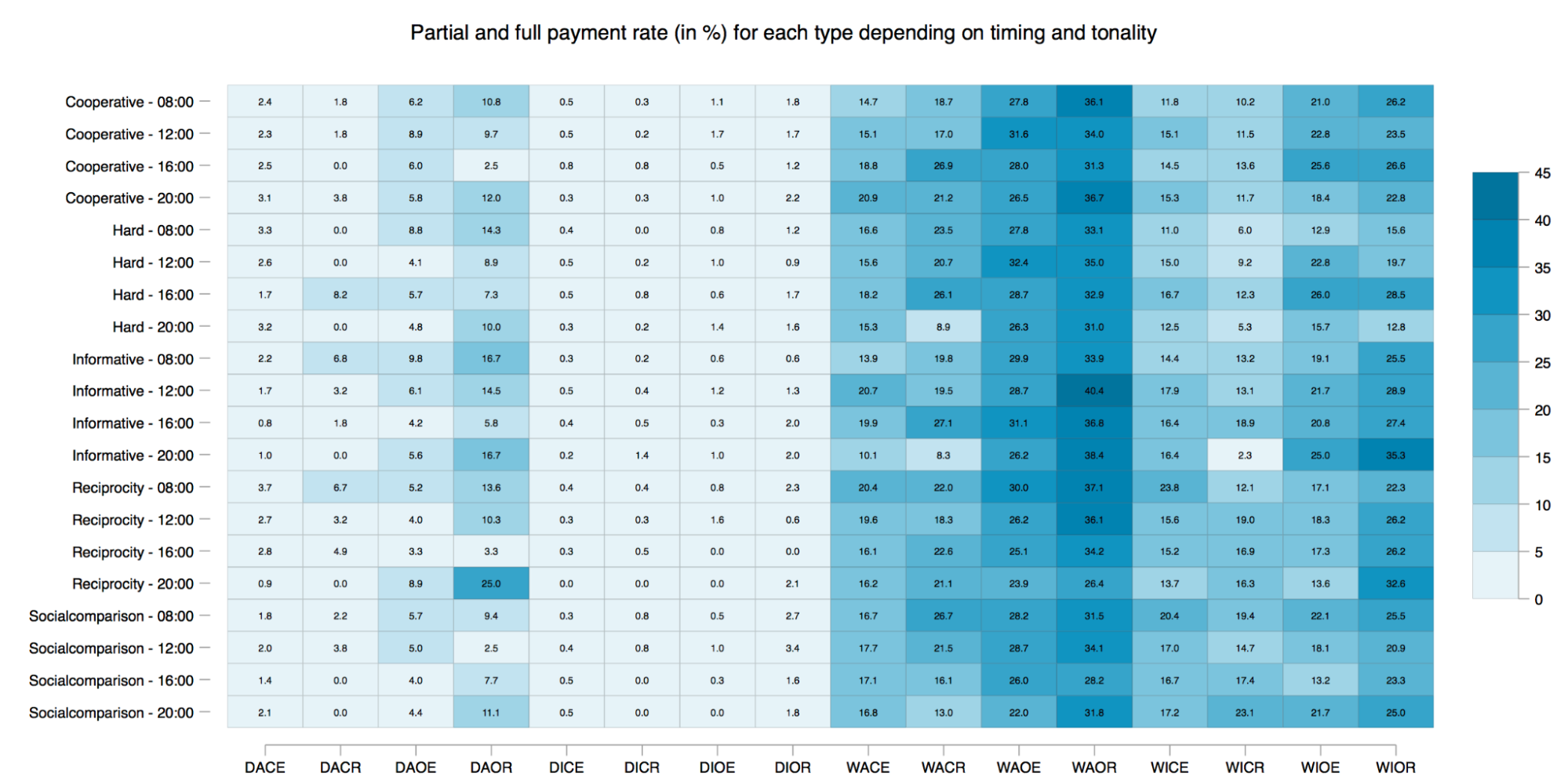}
	\label{fig:payment_timing}
\end{figure}

\end{document}